\documentstyle[12pt,epsfig]{article}

\advance \textheight by \topskip
\newcommand{\ybox}[2]	{
 \begin{center}
 \resizebox{!}{#1\textheight}
{\includegraphics{#2.eps}}
 \end{center}		}

\begin{document}

\begin{center}
{\LARGE \bf The Effect of Non Gaussian Errors on the Determination of Steeply\\
Falling Spectra}\\

\end{center}

\begin{center}
{C.O. Escobar$^1$, L. G. dos Santos$^1$, and R.A.~V\'azquez$^2$ \\}
{\it $^1$ Instituto de Fisica Gleb Wataghim, Unicamp \\}
{\it $^2$ Departamento de F\'{\i}sica de Part\'{\i}culas,\\
Universidade de Santiago, 15706 Santiago de Compostela, Spain\\}
\end{center}

\begin{abstract}
The determination of a steeply falling energy spectrum from rare events with
non-gaussian measurement errors is a delicate matter. The final shape
of the spectrum may be severely distorted as a consequence of
non-gaussian tails in the energy resolution of the experiment. We illustrate
this effect with the recent experimental efforts to determine the 
ultra-high energy extreme of the cosmic ray spectrum.
\end{abstract}

\section{Introduction}
\label{intro}

We will study the determination of a steeply falling spectrum when
the energy measurement errors show long, asymmetrical, non-gaussian tails.
We will see that for a steep spectrum the effect
of long tail error functions can be dramatic: spectrum slopes, cutoff or other
features could be changed or even erased, depending on the degree of deviation
of the error function (energy resolution) from a gaussian distribution.

This problem, well known to nuclear spectroscopists working 
near the end point of a spectrum \cite{tritium} is, however, often ignored
in areas such as Cosmic Ray physics
and we therefore choose to illustrate it by examining the spectrum of Ultra 
High Energy Cosmic Rays (UHECR) as determined by ground arrays, addressing  
the crucial question of the existence or absence of a cutoff in the spectrum, 
known as the Greisen-Zatsepin-Kuzmin (GZK)cutoff \cite{watsonnagano}.

\section{Spectrum reconstruction with non-gaussian errors}
\label{reconst}

We will consider four cases for the reconstruction error function. 
These cases illustrate the different possibilities. 

Let's start with the simplest, well-known, case of a gaussian error 
reconstruction function. Assume that events of energy $E'$ 
are reconstructed with energy $E$ with a probability given by
\begin{equation}
P(E,E') = N \exp (\frac{-(E-E')^2}{2 \sigma^2}),
\end{equation}
where $\sigma$ is constant.
Then, the effect on an initial power law spectrum, $\phi = A E^{-\gamma}$, 
will be simply given by the convolution of $\phi$ with the error function
\begin{equation}
\tilde \phi(E) = \int_0^\infty dE' P(E,E') \phi(E').
\end{equation}
In this case the integral can be carried out analytically but is not very 
illuminating. For high energy $E$ it can be expanded to give 
\begin{equation}
\tilde \phi(E) = A E^{-\gamma} \left( 1+ \gamma (\gamma-1) \frac{\sigma^2}
{2 E^2} + \ldots \right).
\label{gaussR}
\end{equation}
At large energies, the effect of the reconstruction
energy on the reconstructed flux is very small. This is what is expected 
for gaussian errors and therefore one can safely neglect this effect. The same
occurs for any other initial spectrum. The reconstructed spectrum will differ
from the original one by a function which goes to zero rapidly with increasing
energy.

Now let's consider the log-gaussian distribution
\begin{equation}
P(E,E') = N \exp ( - \frac{\log(E/E')^2}{2 \Delta \xi^2}),
\end{equation}
where $N$ is a normalization constant and $\Delta \xi $ is the standard 
deviation. Then the convolution of $P$ with the power like initial flux is 
\begin{equation}
\tilde \phi(E) = \int_0^\infty dE' \phi(E') P(E,E') = A E^{-\gamma} \exp
( \frac{\Delta \xi^2}{2} (\gamma^2-1 )),
\end{equation}
where we are assuming that $\Delta \xi$ is independent of energy.
We see therefore that for a log-gaussian error with a constant standard 
deviation 
the effect of the convolution is to change the normalization. This is rather
different from the result of Eq.\ref{gaussR}, where the change was negligible 
at large enough energies. Here the flux increases (for $\gamma > 1$) by a 
constant factor. Notice also that the enhancement factor depends strongly on
the initial spectrum index, $\gamma$.
If we put $\gamma= 3$ and $\Delta \xi \sim 15 $\% then $\tilde
\phi /\phi \sim 1.09$, for $\Delta \xi \sim 30 $\% then $\tilde
\phi /\phi \sim 1.43$. Alternatively, for $\gamma=3$, we get 
a factor two enhancement for $\Delta \xi \sim 0.4$.

In order to see what is the effect of very long tails on the reconstruction
of events, let us consider an error function with a long tail
\begin{equation}
P(E,E') = N (\frac{E}{E'})^\alpha (1+ \frac{E}{E'} )^{-\alpha - \beta},
\end{equation}
where $N$ is a normalization and $\alpha$ and $\beta$ are constants.
The energy is reconstructed with a power law distribution of slope $\alpha$
and $-\beta$ for small and large reconstructed energies.
The convolution with the original spectrum gives as before
\begin{equation}
\tilde \phi(E) = A K(\alpha,\beta,\gamma) E^{-\gamma},
\end{equation}
where $K$ is a constant which depends on the indexes $\alpha, \beta, \gamma$. 
As in the log-gaussian case the normalization changes but now the 
effect can be dramatic, depending on the values of $\alpha$ and $\beta$.
For $\alpha= 4$ and $\beta = 7$ we find $\tilde \phi/\phi \sim 1.5$.
The log-gaussian distribution can be thought as the limit when $\alpha$ and 
$\beta$ are large.

In both cases the normalization is
strongly affected. This is easily understood, the long tails on the
error functions favor the more numerous low energy events to be
reconstructed at higher energy, giving a higher flux.
In all these cases
we are assuming that there are no systematical energy shifts: the average 
reconstructed energy for a fixed initial energy is assumed to be equal to the 
initial energy. The existence of systematic errors in the reconstruction of 
energy would worsen considerably our results since in this case the change
on the normalization of the spectrum will be linear in 
$\Delta \xi_{\rm syst}$ rather than quadratic. Also we are assuming that the 
energy reconstruction function has no strong energy dependence.

Our last example of an error function with long tails is the Moyal distribution
which constitutes an approximation to the Landau distribution describing the
fluctuations in energy losses of an ionizing particle passing through matter 
\cite{moyal}.
\begin{equation}
P(E,E') = N \exp(-\frac{1}{2} (\frac{(E-E')}{\sigma} + e^{-\frac{(E-E')}{\sigma}})).
\end{equation}
We will return to this distribution at the end of the paper when we look in 
detail at the problem of the GZK cutoff in the UHECR spectrum.
Now let's consider an initial spectrum with some feature. In order to keep the 
analysis simple, let's assume a spectrum with a exponential cutoff. 
\begin{equation}
\phi(E) = A E^{-\gamma} \exp(-\frac{E}{E_c}),
\end{equation}
where $E_c$ is the cutoff energy. The effect of a gaussian error would be
again the original flux times a function going rapidly to 1 with energy, {\it
i.e.} the original flux is not modified. 
However for a log-gaussian error or 
for a power law error we will have strong modification of the flux. 
In both cases at low energy ($E \ll E_c$ ) we will have the previous result, 
the reconstructed 
flux is modified by a constant factor. At high energies, $E \gg E_c$,
the effect is different. In the log-gaussian error function we have
\begin{equation}
\tilde \phi(E) = A E^{-\gamma} \exp(\Delta \xi^2/2 (\gamma^2-1)) 
e^{-E/\tilde E_c},
\end{equation}
where now $\tilde E_c = E_c \exp(\gamma \Delta \xi^2)$ is the new, smeared, 
cutoff energy. Again, this effect
can be large due to the non linear dependence on $\gamma$ and $\Delta \xi$; 
for $\Delta \xi = 0.3$ and $\gamma = 3$ $\tilde E_c = 1.3 E_c$, for
$\Delta \xi = 0.4$ $\tilde E_c = 1.62 E_c$.

In the case of a power law error function the effect is even more dramatic. At 
high energy, $E \gg E_c$, the convolution of the error function with the 
exponential cutoff flux is 
\begin{equation}
\tilde \phi(E) =  A E^{-\beta} E_c^{\beta -\gamma} K'(\alpha,\beta,
\gamma),
\end{equation}
{\it i.e.} the cutoff is completely washed out and replaced by a 
spectrum with a constant slope. The new slope is solely dependent on the
error function and not on the original spectrum. 

\begin{figure} 
\ybox{0.3}{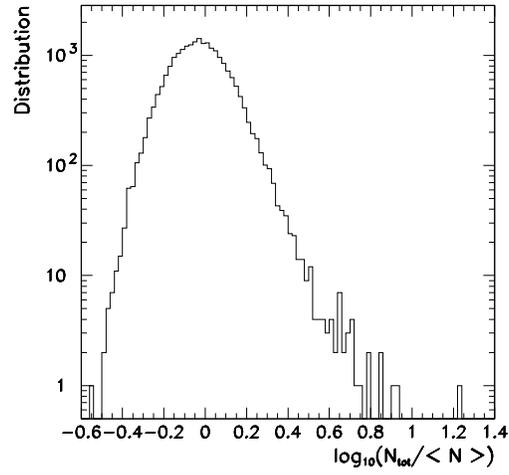}
\caption{Distribution of the total number of charged particles at ground
level for protons of $E=1 $ EeV generated using Aires with Sibyll hadronic 
generator.}
\label{fig1}
\end{figure}
\begin{figure}
\ybox{0.3}{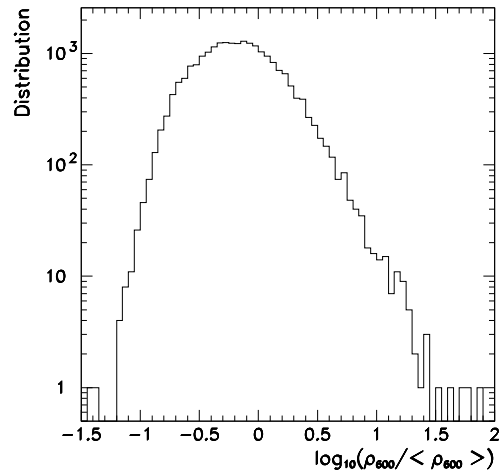}
\caption{Distribution of $\rho(600)$ at ground
for protons of $E=1 $ EeV generated using Aires with Sibyll hadronic 
generator.}
\label{fig2}
\end{figure}

We want now to argue that the non-gaussian error functions given above are 
realistic. There are several reasons why one expects long tails in the
reconstruction of events in cosmic ray experiments. 
In ground array experiments the energy is estimated by measuring the density
of particles in the shower. Usually parameters such as $\rho(600)$, the 
density of particles at $600$ m from the core, are used. It is found
that the energy of the shower scales as  $E \propto \rho(600)^\kappa$ and 
$\kappa \sim 1$. But this quantity
has non-gaussian uncertainties. First, to calculate $\rho(600)$ 
(or any similar parameter) the core position and the arrival direction
of the shower are needed. Given an arrival direction the $\rho(600) $ is 
corrected to ''zero degrees'' arrival direction by an equation like
\begin{equation}
\rho(600)_0 = \rho(600) \exp(s_0(\sec(\theta)-1)),
\end{equation}
where $\theta$ is the zenith angle and $s_0$ is a constant. Even if the
arrival direction is reconstructed with a gaussian probability function
the resulting $\rho(600)_0$ is not. However, this effect is, we believe, small
at all but the largest zenith angles. The effect on the core reconstruction
is more important. The density of particles at a distance $r$ from the
shower core can be 
parameterized by the NKG lateral distribution function
\begin{equation}
\rho(r) = K r^\eta (1+r/R)^{\eta'},
\end{equation}
where $K,R,\eta,\eta'$ are constants. Then, the error in the density due
to the error in the determination of the core position would be generally 
non-gaussian. 

Finally, any indirect measurement of cosmic rays based on shower
development is subjected to shower fluctuations. It is well known that 
the total number of particles at fixed depth have large non-gaussian 
fluctuations, the total number of particles r.m.s scales as the energy.
The same occurs for $\rho(600)$, since it is related to the total number
of particles. In figure \ref{fig1}
we can see the distribution of the total number of particles at ground
level in a shower of fixed energy. We can appreciate that the distribution
of the number of particles is far from a gaussian and has long tails. 
In figure \ref{fig2} we show the distribution of the density at 600 meters
for showers of fixed energy. More than 20000 protons showers of energy 
$10^{18}$ were simulated with the Aires Monte Carlo code \cite{Aires}.
As before the distribution
is non-gaussian with tails extending to more than 1 order of magnitude above
the average.

This was known for a long time but rarely used in the analysis of cosmic
ray physics \cite{roma}.
Apparently, it is a generic phenomena in probability theory and does not 
contradict the central limit theorem \cite{Tribelsky}. 

\section{The GZK Cutoff}
\label{algo.sec}

Our results are not directly applicable to any real experiment. Rather, 
experiments should ascertain what are their error reconstruction functions 
and take them into account in the calculation of the spectrum. Also,
events with large fluctuations could be cut off by other methods, improving
therefore the energy reconstruction and avoiding these unwanted effects.
Particularly important would be to get rid of any power like tail in the
reconstruction error functions, since they have the most devastating effects
on the spectrum reconstruction. 
On the other hand this effect could, at least in part, explain the current 
experimental situation where the normalization of the cosmic ray flux for 
different experiments is different and the 
existence of the GZK cutoff is controversial. Values of $\Delta \xi \sim 
30 - 40 $ \% are usual in cosmic ray physics and as said before with such
values there are important effects on the shape and normalization of the
resulting spectra.
\newpage
\begin{figure}
\ybox{0.3}{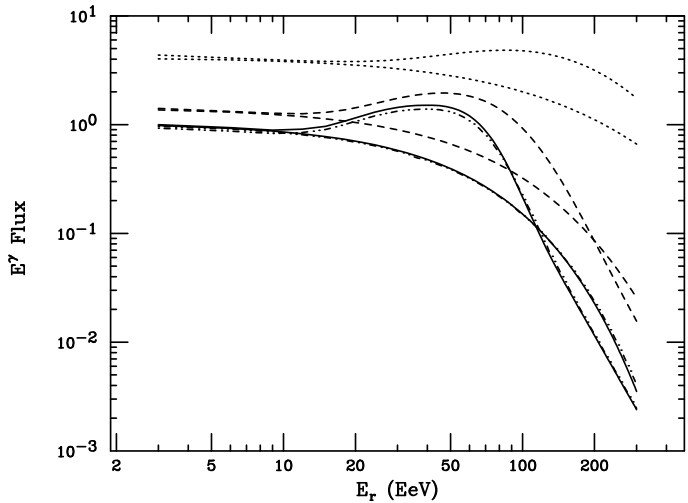}
\caption{Expected flux for log-gaussian energy reconstruction error function
with $\Delta \xi = 0.1,0.3,0.6$ (continuous, dashed, and dotted lines) for
a flux with exponential cut-off at $E=50$ EeV and for a cosmological 
distribution of sources.}
\label{fig3}
\end{figure}

The current experimental situation is aggravated by the low statistics in the
high energy part of the spectrum. The Agasa experiment has measured $58$
\cite{AgasaGZK} events above $4 \times 10^{19}$ eV, of them 8 are above
$10^{20}$ eV. Using a realistic spectrum obtained from an uniform cosmological
distribution of sources \cite{Teshima} and using a log-gaussian error 
function with $\Delta \xi=0.4$ we obtain that for a total of 58 events above
$4 \times 10^{19}$ eV, the probability of having 8 or more events above 
$10^{20}$ eV is 2\%. The average number of expected events would be 3.
In figure \ref{fig3} the convolution of different spectrum with energy error
functions is shown. In the case of a $\Delta \xi =0.3$ and the cosmological 
distribution of sources of Teshima and Yoshida \cite{Teshima} one can see that
at $E=200$ EeV only a 0.1 reduction on the flux is expected.

The number of AGASA events above the GZK cutoff seems to establish its 
absence. However, if we introduce non-gaussian errors, we stress once again, 
the observation of such events becomes compatible with the GZK cutoff.    
To illustrate quantitatively the possibility of not seeing the GZK cutoff 
because of non-gaussian errors in the energy determination we consider the 
Moyal distribution of equation 4 (with approximately the same half-width of a 
gaussian distribution  of a given variance, as seen in fig. 4) to each energy 
sampled from the AGASA spectrum with a exponential cutoff, that is a power 
law spectrum with an index of 2.78 multiplied by an exponentially falling 
spectrum starting at $4 \times 10^{19}$ eV.
\begin{figure}
\begin{center}
\mbox{\epsfig{figure=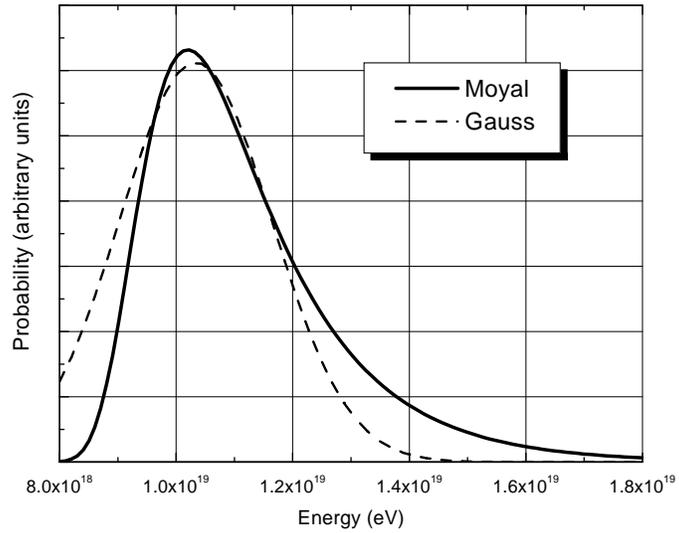,width=7.0cm,angle=-90}}
\caption{The Moyal distribution of Eq. 4 compared to a gaussian distribution
having approximately the same half width}
\label{fig4}
\end{center}
\end{figure}
These errors alter the shape of the original spectrum and the expected number 
of events above the GZK cutoff . We summarize the results of this analysis in 
fig. 5. 
\begin{figure}
\begin{center}
\mbox{\epsfig{figure=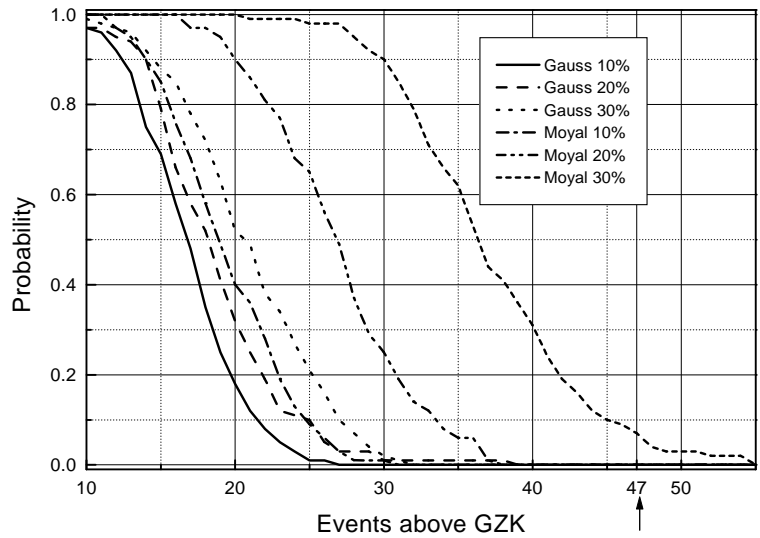,width=7.0cm,angle=-90}}
\caption{Probability of obtaining events above the GZK cutoff for different 
resolutions (error functions)}
\label{fig5}
\end{center}
\end{figure}
We can compare the probability of finding events above GZK for measurements affected 
by gaussian errors (10\%, 20\%, and 30\%) with measurements affected by Moyal-distributed 
errors (with the same half-width of gaussians with standard deviation of 10\%, 20\%, 
and 30\% of energy). We see that considering only gaussian errors we cannot explain the 
observed number of events above GZK by AGASA 
(indicated by an arrow in figure 5) and would be led to conclude that there 
is a cutoff in the UHECR spectrum. However, for Moyal-distributed errors (30\% of 
energy) we have an appreciable probability of finding more than 47 events above the 
GZK cutoff. We have shown in this paper that the experimental resolution curve
when deviating from a gaussian distribution severely impacts the shape and
normalization of a steeply falling spectra near its end point. Spectral 
features such as a cutoff may disappear from the reconstructed spectrum due to 
the effect of long tails in the resolution(error)function of the experiment.

{\bf Acknowledgments:} We thank discussions with M\'aximo Ave and Enrique Zas.
One of us (RAV) thanks the Instituto de Fisica Gleb Wataghim for kind 
invitation when this work took place and to FAPESP,
Xunta de Galicia (XUGA-20604A98), and to the ''Ram\'on y Cajal'' fellowship 
program for financial support. LGS and COE thank FAPESP for its financial 
support as well as the Brazilian Research Council-CNPq (COE).

\end{document}